\documentclass[times, 10pt,twocolumn]{article}
\usepackage{latex8}
\usepackage{times}
\usepackage{subfigure}
\usepackage{xspace}
\usepackage{enumerate}
\newtheorem{theorem}{Theorem}
\newtheorem{proposition}{Proposition}
\newtheorem{lemma}{Lemma}
\newtheorem{definition}{Definition}
\newcommand{\pair}{$\ast$-pair}
\newcommand{\openbox}{\leavevmode
  \hbox to.77778em{%
  \hfil\vrule
  \vbox to.675em{\hrule width.6em\vfil\hrule}%
  \vrule\hfil}}

\def\QEDbox{\begingroup\unitlength\p@\linethickness{.4\p@}\framebox(6,6){}\endgroup}

 \usepackage[small,compact]{titlesec}

\usepackage{comment}
\excludecomment{myproof}
\usepackage{makeidx}
\title{An Optimal Linear Time Algorithm for Quasi-Monotonic Segmentation}

\author{Daniel Lemire\\
University of Quebec at Montreal (UQAM)\\ 100 Sherbrooke West \\ Montréal, Qc,
 Canada, H2X 3P2\\ lemire@ondelette.com\\
\and
Martin Brooks, Yuhong Yan\\
National Research Council of Canada\\
1200 Montreal Road\\ Ottawa, ON, Canada, K1A 0R6\\ 
First.Last@nrc.gc.ca\\
}

\usepackage{pslatex}
\usepackage[T1]{fontenc}
\usepackage[latin1]{inputenc}

\usepackage{graphicx}
\usepackage{latexsym}
\usepackage{setspace}
\graphicspath{{../Python}{..}}
\usepackage{ifpdf}
\ifpdf
\usepackage[pdftex,plainpages=false]{hyperref}
\fi
\newcommand{\yuhong}[1]{} 

\newcommand{\cutforlackofspace}[1] {}
\newcommand{\longpaper}[1] {}
\newcommand{\shortpaper}[1] {#1}
\newcommand{\delete}[1]{}
\usepackage{algorithm}
\usepackage[noend]{algorithmic} 
\usepackage{amsfonts}
\begin{document}



\maketitle
\thispagestyle{empty}

\begin{abstract}
Monotonicity is a simple yet significant qualitative characteristic.  We consider the problem of segmenting an array in up to $K$ segments. We want
segments to be as monotonic as possible and to alternate signs. We propose a quality metric for this problem,
present an optimal linear time algorithm based on novel formalism, and compare experimentally its performance to a linear time top-down regression
algorithm. We show that our algorithm is faster and more accurate. Applications include pattern recognition and qualitative modeling.
\end{abstract}

\section{Introduction}

Monotonicity is one of the most natural and important qualitative properties for sequences of data points. 
It is easy to determine where the values are strictly going up or down, but we only want to
identify significant monotonicity. For example, the drop from 2 to 1.9 in the array $0,1,2,1.9, 3,4$
might not be significant and might even be noise-related. 
 The quasi-monotonic segmentation problem is to determine where the data is approximatively increasing or decreasing.
 
 \longpaper{Some segmentation algorithms give a segmentation having no more than $K$ segments while attempting to minimize the error $\epsilon$; other algorithms attempt to minimize the number of segments ($K$) given an upper bound on the error $\epsilon$. We are concerned with the first type of algorithm in this paper. }
 
\longpaper{
Using dynamic programming or other approaches, most segmentation problems can be solved in time $O(n^2)$. Other solutions to this problem using machine learning to classify the pairs of data points~\cite{kn:SucECML01} are even less favorable since they have higher complexity. However, it is common for sequence of data points to be massive and segmentation algorithms have to have complexity close to $O(n)$ to be competitive. 
 While approximate linear regression segmentation algorithms can be $O(n)$, we show that using a linear regression error to segment according to  monotonicity is not an ideal solution.}

We present a metric for the quasi-monotonic segmentation problem
called the Optimal Monotonic Approximation Function Error (OMAFE); this metric differs
from previously introduced OPMAFE metric~\cite{YLBIJCAI05} since it applies to all segmentations and not just ``extremal'' segmentations. 
We formalize the novel concept of a maximal \pair{} and shows that it can be used to 
define a unique labelling of the extrema leading to an optimal segmentation algorithm.
We also present an optimal linear time algorithm to solve the quasi-monotonic segmentation problem given a segment budget together with an experimental comparison to quantify the benefits of our algorithm. 



\section{Monotonicity Error Metric (OMAFE)}


Suppose $n$ samples noted $F: D=\{x_,\ldots, x_n\} \rightarrow \mathbb{R}$ 
with $x_1 < x_2 < \ldots x_n$.
We define, $F_{|[a,b]}$ as the restriction of $F$ over $D\cap [a,b]$. We seek the best monotonic
(increasing or decreasing) function $f:\mathbb{R}\rightarrow \mathbb{R}$ approximating $F$. Let
$\Omega_{\uparrow}$ (resp. $\Omega_{\downarrow}$) be the set of all monotonic increasing (resp. decreasing)
functions. The \textbf{Optimal Monotonic Approximation Function Error (OMAFE)} is $\min_{f \in \Omega}
\max_{x\in D} \vert f(x) - F(x)\vert$ where $\Omega$ is either $\Omega_{\uparrow}$ or $\Omega_{\downarrow}$.

%
%

The segmentation of a set $D$ is a sequence $S = X_1 ,\dots,  X_K$ of intervals in $\mathbb{R}$ with $[\min
D,\max D] = \bigcup_i X_i $ such that $\max X_i =\min X_ {i+1} \in D$ and $X_i \cap X_j= \emptyset$ for
$j \neq i+1, i, i-1$. Alternatively, we can define a segmentation from the set of points $X_i \cap X_
{i+1}=\{y_{i+1}\}$, $y_1= \min X_1$, and $y_{K+1}=\max X_K$. Given $F: \{x_1,\ldots, x_n\} \rightarrow \mathbb{R}$ and a segmentation, the Optimal Monotonic
Approximation Function Error (OMAFE) of the segmentation is  $\max _i \textrm{OMAFE}(F_{|X_i})$ where
the monotonicity type (increasing or decreasing) of the segment $X_i$ is determined by the sign of $F(\max X_i)-F(\min X_i)$. Whenever $F(\max X_i)=F(\min X_i)$, we say the segment has no direction and the best monotonic approximation is just the flat function having value $(\max F_{|X_i} - \min F_{|X_i})/2$. The error is computed over each interval independently; optimal monotonic approximation functions are not required to agree at $\max X_i =\min X_ {i+1}$. Segmentations should alternate between increasing and decreasing, otherwise sequences such as $0,2,1,0,2$ can be segmented as two increasing segments $0,2,1$ and $1,0,2$: we consider it is natural to aggregate segments with the same monotonicity.

We solve for the best monotonic function as follows. 
If we seek the best monotonic increasing function, we first define
$\overline{f}_{\uparrow} (x) = \max \{ F(y): y \leq x \}$ (the maximum of all previous values) and
$\underline{f}_{\uparrow} (x) = \min \{ F(y): y \geq x \}$ (the minimum of all values to come).
 If we seek the best monotonic decreasing function,
we define $\overline{f}_{\downarrow} (x) =  \max \{ F(y): y \geq x \}$ (the maximum of all values to come) and
$\underline{f}_{\downarrow} (x) =\min \{ F(y): y \leq x \}$ (the minimum of all previous values).
These functions, which can
be computed in linear time, are all we need to solve for the best approximation function as shown by the next theorem
which is a well-known result~\cite{Ubhaya1974}.

\begin{theorem}
Given $F: D=\{x_1,\ldots, x_n\} \rightarrow \mathbb{R}$, a best monotonic increasing approximation function to $F$ is $f_{\uparrow}= (\overline{f}_{\uparrow} + \underline{f}_{\uparrow})/2$ and a best monotonic
decreasing approximation function is $f_{\downarrow}=(\overline{f}_{\downarrow} +
\underline{f}_{\downarrow})/2$. The corresponding error (OMAFE) is $\max_{x\in D} (\vert
\overline{f}_{\uparrow}(x) - \underline{f}_{\uparrow}(x) \vert)/2$  or $\max_{x\in D}
(\vert \overline{f}_{\downarrow}(x) - \underline{f}_{\downarrow}(x) \vert)/2$ respectively.
\end{theorem}

\longpaper{The implementation of the algorithm suggested by the theorem is straight-forward.
 Given a segmentation, we can compute the OMAFE in $O(n)$ time
using at most two passes.}

\section{A Scale-Based Algorithm for Quasi-Monotonic Segmentation}
\cutforlackofspace{
 The following propositions describe conditions for an optimal segmentation of $F$ if only alternating segmentations are allowed. It sets the stage for a scale-based approach. The next proposition is not used later on but is given as a reference.
 \begin{proposition}\label{stupidresult}
 A segmentation $S=[y_1=x_1,y_2],[y_2,y_3],\ldots,[y_{K},y_{K+1}=x_n]$ of $F: D=\{x_1,\ldots,x_n\} \rightarrow
 \mathbb{R}$ with alternating monotonicity has a minimal number of alternating segments $K$ for a given  $\epsilon$  if for all $i\in \{1,\ldots,K\}$, $[y_i,y_{i+1}]$ has OMAFE less than
 $\epsilon$ (with direction given by $F(y_{i+1})-F(y_i)$) and  $\vert F(y_{i+1})-F(y_i) \vert  \geq 2 \epsilon$.
 \end{proposition}
 \begin{myproof}
  Because all segments have OMAFE bounded by $\epsilon$, the segmentation has OMAFE less than $\epsilon$.
   The fact that $K$ is minimal follows
  from  $\vert F(y_{i+1})-F(y_i) \vert  \geq 2 \epsilon$. 
  Suppose you have a better segmentation with fewer segments. Consider $[y_i,y_{i+1}]$: either the interval is partly contained in a same monotonicity segments, or is contained in a different monotonicity segment.  We have that it cannot be contained in a  different monotonicity segment because  $\vert F(y_{i+1})-F(y_i) \vert  \geq 2 \epsilon$ would imply that the OMAFE is at least $\epsilon$ over this interval. One the other hand, if all intervals intersect a new segment with same monotonicity, then the new alternating segmentation has the same number of segments.
 \end{myproof}}

 We use the following proposition to prove that the segmentations we generate are optimal (see Theorem~\ref{thm:validlabelling}).
 \begin{proposition}\label{stupidresult2}
 A segmentation $y_1,\ldots,y_{K+1}$ of $F: D=\{x_1,\ldots,x_n\} \rightarrow
 \mathbb{R}$ with alternating monotonicity has a minimal OMAFE $\epsilon$ for a number of alternating segments $K$ if 
\vspace*{-1mm}
 \begin{enumerate}[A.]
\itemsep=0pt\topsep=0pt\partopsep=0pt
\parskip=0pt\parsep=0pt
 \item $F(y_i)=\max F([y_{i-1},y_{i+1}])$ or $F(y_i)=\min F([y_{i-1},y_{i+1}])$ for $i=2,\ldots,K$;
 \item in all intervals $[y_i,y_{i+1}]$  for $i=1,\ldots,K$, there exists $z_1, z_2$ such that $\vert F(z_2)-F(z_1)\vert > 2\epsilon$.
\end{enumerate}
 \end{proposition}
 \longpaper{\begin{myproof}
 Let the original segmentation be the intervals $S_1,\ldots,S_K$ and consider a new segmentation with intervals $T_1, \ldots, T_K$. Assume that the new segmentation has lower error (as given by OMAFE).
 Let $S_i=[y_i, y_{i+1}]$ and $T_i = [y'_i , y'_{i+1}]$.

If any segment $T_m$ contains a segment $S_j$, then the existence of $z_1, z_2$ in $[y_j,y_{j+1}]$ such that $|F(z_2)-F(z_1)| > 2\epsilon$ and $\textrm{OMAFE}(T_m) \le \epsilon$ implies that $T_m$ and $S_j$ have the same monotonicity.

We show that each pair of intervals $S_i$, $T_i$  has nonempty intersection. Suppose not, and let $i$ be the smallest index such that $S_i  \subset  T_{i-1}$.  Since $S_i$ and $T_{i-1}$ have the same monotonicity, for each $j < i$, $S_j$ and $T_j$ have opposite monotonicity. Now consider the $i-1$ intervals $T_1, \ldots, T_{i-1}$  and the $i$ points $y_1, \ldots, y_i$. At least one interval contains two consecutive points; choose the largest  $j < i$ such that $T_j$ contains $y_j, y_{j+1}$.  But then $S_j \subset T_j$, contradicting at least one of the assumptions  $|F(z_2) - F(z_1)| > 2\epsilon$ for $z_1,z_2\in S_i$ and $\textrm{OMAFE}(T_j) \le \epsilon$.

It now follows that each pair of intervals $S_i, T_i$  has the same monotonicity.

Since $\textrm{OMAFE}(T) < \textrm{OMAFE}(S)$, we can choose an index $j$ such that $\textrm{OMAFE}(T_j) < \textrm{OMAFE}(S_j)$. We show that there exists another index $p$ such that $\textrm{OMAFE}(T_p) \ge \textrm{OMAFE}(S_j)$, thus contradicting $\textrm{OMAFE}(T) < \textrm{OMAFE}(S)$.  Suppose $S_j$ is increasing; the proof is similar for the opposite case. Then there exist $x <z \in S_j$ such that $F(x) - F(z) = 2 \times OMAFE(S_j)$. From $\textrm{OMAFE}(T_j) < \textrm{OMAFE}(S_j)$ it follows that at least one of $x$ or $z$ lies in $S_j - T_j$, and hence $F(x) - F(y_j) \ge 2 \times \textrm{OMAFE}(S_j)$ or $F(y_{j+1}) - F(z) \ge 2 \times \textrm{OMAFE}(S_j)$. Thus $\textrm{OMAFE}(T_p) \ge \textrm{OMAFE}(S_j)$ for either $p = j-1$ or $p = j+1$. 
  \end{myproof}}

For simplicity, we assume $F$ has no consecutive equal values, i.e.  $F(x_i) \ne F(x_{i+1})$ for $i = 1,\dots, n-1$; our algorithms assume all but one of consecutive equal values values have been removed.
We say  $x_i$ is a maximum if $i \ne 1$ implies $F(x_i) > F(x_{i-1})$ and if $i \ne n$ implies $F(x_i) > F(x_{i+1})$. Minima are defined similarly. 

Our mathematical approach is based on the concept of $\delta$-pair\longpaper{~\cite{brooks94} (see Fig.~\ref{deltapair})}:
\begin{definition}
The tuple $x,y$ ($x<y \in D$) is a $\delta$-pair (or a pair of scale $\delta$) for $F$ if $|F(y)-F(x)| \geq \delta$ and for
all $z \in D$, $x<z<y$ implies $|F(z)-F(x)|< \delta$ and $|F(y)-F(z)|< \delta$. A $\delta$-pair's \textit{direction} is \textit{increasing} or \textit{decreasing}
according to whether $F(y)>F(x)$ or $F(y)<F(x)$.
\end{definition}
  $\delta$-Pairs having opposite directions cannot overlap but they may share an end~point.
$\delta$-Pairs of the same direction may overlap, but may not be nested. We use the term ``\pair{}'' to
indicate a $\delta$-pair having an unspecified $\delta$. We say that a \pair{} is significant at scale $\delta$ if it is of scale $\delta'$ for $\delta' \geq \delta$.

\longpaper{\begin{figure}
\center{%
\includegraphics[width=0.8\columnwidth]{../deltapair}
}
\caption{\label{deltapair} A $\delta$-pair.  }
\end{figure}}

 We define $\delta$-monotonicity as follows:
 \begin{definition}
 Let $X$ be an interval, $F$ is $\delta$-\textit{monotonic} on $X$ if all
 $\delta$-pairs in $X$ have the same direction; $F$ is \textit{strictly} $\delta$-monotonic when there exists at least one such $\delta$-pair. In this case:
\vspace*{-1mm}
 \begin{itemize}
\itemsep=0pt\topsep=0pt\partopsep=0pt
\parskip=0pt\parsep=0pt
 \item $F$ is $\delta$-\textit{increasing} on $X$ if $X$ contains an increasing
 $\delta$-pair.
 \item $F$ is $\delta$-\textit{decreasing} on $X$ if $X$ contains a decreasing
 $\delta$-pair.
 \end{itemize}
  \end{definition}

A $\delta$-monotonic interval $X$ satisfies $\textrm{OMAFE}(X)<\delta/2$. 
We say that a \pair{} $x,y$ is \textbf{maximal} if whenever $z_1,z_2$ is a \pair{} of a larger scale in the same direction containing $x,y$, then there exists a \pair{} $w_1,w_2$ of an opposite direction contained in $z_1,z_2$ and containing $x,y$.  For example, the sequence $1,3,2,4$ has 2 maximal \pair{}s: $1,4$ and $3,2$. 
  Maximal \pair{}s of opposite direction may share a common point, whereas maximal \pair{}s of the same direction may not. Maximal \pair{}s cannot overlap, meaning that it cannot be the case that exactly one end point of a maximal \pair{} lies strictly between the end points of another maximal  \pair{}; either neither point lies strictly between or both do. In the case that both do, we say that the one maximal \pair{} properly contains the other. All \pair{}s must be contained in a maximal \pair{}.

 \begin{lemma}
 The smallest maximal \pair{} containing a \pair{} must be of the same direction.
 \end{lemma}
 
\begin{myproof} Suppose a \pair{} is immediately contained in a maximal \pair{} $W$. Suppose $W$ is not
in the same direction, then within $W$, seek the largest \pair{} in the
same direction as $P$ and containing $P$, then it must be a maximal \pair{} in $D$ since maximal
\pair{}s of different directions cannot overlap.\end{myproof}

The first and second point of a maximal \pair{} are extrema and the reverse is true as well as shown by the next lemma. 

\begin{lemma}\label{prop:either}
Every extremum is either the first or second point of a maximal \pair{}.
\end{lemma}
\begin{myproof}
The case $x=x_1$ or $x=x_n$ follows by inspection. Otherwise, $x$ is the end point of a left and a right \pair{}. Each \pair{} must immediately belong to a maximal \pair{} of same direction: a \pair{} $P$ is contained in a maximal \pair{} $M$ of same direction and there is no maximal \pair{} $M'$ of opposite direction such that $P\subset M'\subset M$. Let $M^l$ and $M^r$ be the maximal \pair{}s immediately containing the left and right \pair{} of $x$. Suppose neither $M^l$ and $M^r$ have $x$ as a end point. Suppose $M^l \subset M^r$, then the right \pair{} is not immediately contained in $M^r$, a contradiction.  The result follows by symmetry.
\end{myproof}


Our approach is to label each extremum in $F$ with a scale parameter $\delta$ saying that this extremum is ``significant'' at scale $\delta$ and below.
%
Our intuition is that by picking extrema at scale $\delta$, we should have a segmentation having error less than $\delta/2$.

\begin{definition}\label{def:optscalelabel}The scale labelling of an extremum $x$ is the maximum of the scales of the maximal \pair{}s for which it is an end point.
\end{definition}

For example, given the sequence $1,3,2,4$ with  2 maximal \pair{}s ($1,4$ and $3,2$), we would give the following labels in order $3,1,1,3$.

\begin{definition}Given $\delta > 0$, a {\it maximal alternating sequence of $\delta$-extrema} 
$Y = y_1 \dots y_{K+1}$ 
is a sequence of extrema each having scale label at least $\delta$, having alternating types (maximum/minimum), and such that there exists no
sequence properly containing $Y$ having these same properties. From $Y$ we define a
{\it maximal alternating $\delta$-segmentation} of $D$ by segmenting at the points $x_1, y_2 \dots y_K, x_n$. 
\end{definition}

\begin{theorem}\label{thm:validlabelling}
Given $\delta > 0$, let $P = S_1 \dots S_K$ be a maximal 
alternating $\delta$-segmentation derived from maximal alternating sequence $y_1 \dots y_{K+1}$ of $\delta$-extrema.
Then any alternating segmentation $Q$ having OMAFE($Q$) $<$ OMAFE($P$) has at least $K+1$ segments.
\end{theorem}

\begin{myproof} We show that conditions A and B of Proposition~\ref{stupidresult2} are satisfied with $\epsilon =$ OMAFE($P$).

First we show that each segment $S_i$ is $\delta$-monotone; from this we conclude that $\textrm{OMAFE}(P) < \delta/2$.
Intervals $[x_1,  y_1]$ and $[y_K, x_n]$ contain no maximal \pair{}s of scale $\delta$ or larger, and
therefore contain no \pair{}s of scale $\delta$ or larger. Similarly, no $[y_i, y_{i+1}]$ contains an opposite-direction significant
\pair{}.

Condition A: Follows from $\delta$-monotonicity of each $S_i$ and maximal \pair{}s not overlapping.

Condition B: We show that $\vert F(y_{i+1})-F(y_i) \vert  \geq \delta > 2 \times \textrm{OMAFE}(P)$. If $i=1$, then $y_i$ must begin an maximal \pair{}, and the maximal \pair{} must end with $y_{i+1}$ since maximal \pair{}s cannot overlap. The case $i+1=k$ is similar. Otherwise, since maximal \pair{}s cannot overlap, each $y_i, y_{i+1}$ is either  a maximal \pair{} of scale $\delta$ or larger or there exist indices $j$ and $k$, $j < i$  and $k > i+1$ such that $y_j , y_i$ is a maximal \pair{} of scale at least $\delta$, and $y_{i+1}, y_k$ is a maximal \pair{} of scale at least $\delta$. These two maximal \pair{}s have the same direction, and that this is opposite to the direction of $[y_i  y_{i+1}]$. 
Now suppose  $\vert F(y_i) - F(y_{i+1})| < \delta$. Then $y_j, y_k$ is a \pair{} properly containing $y_j$, $y_i$ and $y_{i+1}$, $y_k$. 
But neither $y_j , y_i$ nor $y_{i+1}, y_k$ can be properly contained in a \pair{} of opposite direction lying within
$y_j, y_k$, thus contradicting their maximality and proving the claim. 
\end{myproof}

Sequences of extrema labelled at least $\delta$ are generally not maximal alternating. For example the sequence 
$0,10,9,10,0$ is scale labelled $10,10,1,10,10$.
However, a simple relabelling of certain extrema can make them maximal alternating. Consider two same-sense extrema $z_1 < z_2$
such that lying between them there exists no extremum having scale at least as large as the minimum of the two extrema's scales. We must have $F(z_1) = F(z_2)$, since otherwise the point upon which $F$
has the lesser value could not be the endpoint of a maximal \pair{}.
This is the only situation which causes choice when constructing a maximal alternating sequence of $\delta$-extrema.
To eliminate this choice, replace the scale label on $z_1$ with the largest scale of the opposite-sense extrema lying between them.
\longpaper{In the next section, Algorithm~\ref{algo:computedelta} incorporates this re-labelling making Algorithm~\ref{algo:sortdelta} simple and efficient. }

\cutforlackofspace{

In light of Proposition~\ref{stupidresult2}, the next proposition says that the significant extrema gives an optimal segmentation except when there are successive maxima or minima as in this sequence  $0,10,9,10,-1$ at a scale strictly greater than 1.

\begin{proposition}\label{prop:validlabelling}
If for a given $\delta$, all significant
extrema of $F: D=\{x_1,\ldots,x_n\} \rightarrow \mathbb{R}$ at scale $\delta$, $\{y_1,\ldots, y_K\}\subset D$, are
such that there are no successive maxima or successive minima then
\vspace*{-1mm}
\begin{itemize}
\itemsep=0pt\topsep=0pt\partopsep=0pt
\parskip=0pt\parsep=0pt
\item $F$ is $\delta$-monotonic on $[y_i, y_{i+1}]$ for all $i$;
\item $\vert F(y_{i+1})-F(y_i) \vert  \geq \delta$ for all $i$;
\item $F(y_i)= \max F([y_{i-1},y_{i+1}])$ or $F(y_i)= \min F([y_{i-1},y_{i+1}])$ for all $i$; 
\item there are no \pair{} of scale $\delta$ or more in $[x_1,y_1]$ or in $[y_{K-1}, x_n]$.
\end{itemize}
\end{proposition}
\begin{myproof}
Suppose that $[y_i,y_{i+1}]$ is not $\delta$-monotonic. For simplicity, suppose $y_i$ is a maximum and $y_{i+1}$ is a minimum. We have that $[y_i,y_{i+1}]$ must
contain an increasing $\delta'$-pair for $\delta'\geq\delta$ (otherwise, it is $\delta$-monotonic) which implies we omitted a significant minimum followed by a maximum inside $[y_i,y_{i+1}]$, a contradiction.

Next we show that $\vert F(y_{i+1})-F(y_i) \vert  \geq \delta$. If $i=1$, then $y_i$ must begin an maximal \pair{}, and the maximal \pair{} must end with $y_{i+1}$ since maximal \pair{}s cannot overlap. The case $i+1=k$ is similar. Otherwise, since maximal \pair{}s cannot overlap, each $y_i, y_{i+1}$ is either  a maximal \pair{} of scale $\delta$ or larger or there exist indices $j$ and $k$, $j < i$  and $k > i+1$ such that $y_j , y_i$ is a maximal \pair{} of scale at least $\delta$, and $y_{i+1}, y_k$ is a maximal \pair{} of scale at least $\delta$. 
These two maximal \pair{}s have the same direction, and that this is opposite to the direction of $[y_i  y_{i+1}]$. 
Now suppose  $\vert F(y_i) - F(y_{i+1})| < \delta$. Then $y_j, y_k$ is a \pair{} properly containing $y_j$, $y_i$ and $y_{i+1}$, $y_k$. 
But neither $y_j , y_i$ nor $y_{i+1}, y_k$ can be properly contained in a \pair{} of opposite direction lying within
$y_j, y_k$, thus contradicting their maximality and proving the claim.

Suppose there is a \pair{} of scale at least $\delta$ in $[x_1,y_1]$, then we have that both end~points must be significant extrema contradicting the fact that $y_1$ is the first significant extrema. Other results follow by inspection.
\end{myproof}

 The next proposition shows that when we have successive significant maxima or successive significant minima, then they have the same value.
 
\begin{proposition}\label{prop:repeatedextrema}
For a given $\delta$, if $y,z\in D$ are
successive significant maxima or successive significant minima, then $F(y)=F(z)$.
\end{proposition}
\begin{myproof}
If there is no significant extrema between $y,z\in D$, then there is no significant maximal \pair{} in $[y,z]$. Because $y$ and $z$ are significant. there exists $y'$ and $z'$ such that $y',y$ and $z,z'$ are significant maximal \pair{}s of opposite direction. If $F(y) \neq F(z)$, we can extend either $y',y$ or $z,z'$, a contradiction.
\end{myproof}

For an optimal segmentation, it suffices to prune successive maxima or minima and we call the result of such a pruning a $\delta$-segmentation.

}


\cutforlackofspace{
In light of Proposition~\ref{prop:unique}, we note the valid labelling of an extremum $x$ of $F$ to be $\delta_F(x)$. We will define an \textit{optimal labelling} to be a valid labelling with repeated extrema as a special case.
 
 \begin{definition}Given a function $F$ choose any sequence of functions $F^{(i)}$ with no two extrema have the same value such that $F^{(i)} \rightarrow F$ pointwise over D. If $\Delta_F(x)=\lim_{i\rightarrow \infty} \delta_{F^{(i)}}(x)$ exists for all extrema $x$ of $F$, then $\Delta_F(x)$ is an optimal labelling.\end{definition}

An optimal labelling always exists.
 }
 
\cutforlackofspace{ 
To ensure that the significant extrema form a $\delta$-segmentation, we simply relabel some of the extrema having the same value. Given any two maxima $x_1$ and $x_2$ (resp. minima) at scale $\delta_1$ and $\delta_2$ such that the right maximal \pair{} of $x_1$ is at scale $\delta'<\min (\delta_1, \delta_2)$, then we label $x_1$ with $\delta'$.
  The following result follows from Propositions \ref{stupidresult2} and \ref{prop:validlabelling}.

\begin{theorem}\label{thm:validlabelling}
For a given $\delta$, any $\delta$-segmentation is an optimal segmentation having OMAFE less than $\delta/2$: no segmentation having the same number of alternating segments can have a smaller OMAFE.
\end{theorem}

Hence, given a budget of $K$ segments, and pick $\delta$ as small as possible so that you have no more than $K+1$ significant extrema, and you have an optimal segmentation. Also, should the original data be already piecewise monotonic, picking the right number of monotonic segments $K$ leads to a perfect segmentation. 
}

\subsection{Computing a Scale Labelling Efficiently}

Algorithm~\ref{algo:computedelta} (next page) produces a scale labelling in linear time. Extrema from the original data are visited in order, and they alternate (maxima/minima) since we only pick one of the values when there are  repeated values (such as $1,1,1$).  

The algorithm has a main loop (lines 5 to 12) where it labels extrema as it identifies extremal \pair{}s, and stack the extrema it cannot immediately label.  At all times, the stack (line 3) contains minima and maxima in \textbf{strictly} increasing and decreasing order respectively. Also at all times, the last two extrema at the bottom of the stack are the absolute maximum and absolute minimum (found so far).  Observe that we can only label an extrema as we find new extremal \pair{}s (lines 7, 10, and 14).
\vspace*{-1mm}
\begin{itemize}
\itemsep=0pt\topsep=0pt\partopsep=0pt
\parskip=0pt\parsep=0pt
\item If the stack is empty or contains only one extremum, we simply add the new extremum (line 12).

\item If there are only 2 extrema $z_1,z_2$ in the stack and we found either a new absolute maximum or new absolute minimum ($z_3$), we can pop and label the oldest one ($z_1$) (lines 9, 10, and 11) because the old pair ($z_1,z_2$) forms a maximal \pair{} and thus must be bounded by extrema having at least the same scale while the oldest value ($z_1$) doesn't belong to a larger maximal \pair{}. 
Otherwise, if there are only 2 extrema $z_1,z_2$ in the stack and the new extrema $z_3$ satisfies $z_3 \in (\min(z_1,z_2), \max(z_1,z_2))$, then we add it to the stack since no labelling is possible yet.

\item While the stack contains more than 2 extrema (lines 6, 7 and 8), we consider the last three points on the stack ($s_3,s_2,s_1$) where $s_1$ is the last point added. Let $z$ be the value of the new extrema. If $z \in (\min(s_1,s_2), \max(s_1,s_2))$, then it is simply added to the stack since we cannot yet label any of these points; we exit the while loop. Otherwise, we have a new maximum (resp. minimum) exceeding (resp. lower) or matching the previous one on stack, and hence $s_1,s_2$ is a maximal \pair{}. If $z\neq s_2$, then $s_3, z$ is a maximal \pair{} and thus, $s_2$ cannot be the end of a maximal \pair{} and $s_1$ cannot be the beginning of one, hence both $s_2$ and $s_1$ are labelled. If $z=s_2$ then we have successive maxima or minima and the same labelling as $z\neq s_2$ applies.
\end{itemize}
During the ``unstacking'' (lines 13 and following), we visit a sequence of minima and maxima 
forming increasingly larger maximal \pair{}s.

\begin{algorithm}

 \begin{algorithmic}[1]
 \STATE \textbf{INPUT:} an array $d$ containing the $y$ values indexed from $0$ to $n-1$, repeated consecutive values have been removed
 \STATE \textbf{OUTPUT:} a scale labelling for all extrema
 \STATE $S\leftarrow$ empty stack, First($S$) is the value on top, Second($S$) is the second value
 \STATE \textbf{define} $\delta(d,S)=\vert d_{\textrm{First}(S)}-d_{\textrm{Second}(S)}\vert$
 \FOR {$e$ index of an extremum in $d$, $e$'s are visited in increasing order}
 \WHILE{length($S$) $>2$ and ($e$ is a minimum such that $d_e\leq \textrm{Second}(S)$ or $e$ is a maximum such that $d_e\geq \textrm{Second}(S)$)}
 \STATE label First($S$) and Second($S$) with $\delta(d,S)$
 \STATE pop stack $S$ twice
 \ENDWHILE
 \IF{length($S$) is 2 and ($e$ is a minimum such that $d_e\leq \textrm{Second}(S)$ or $e$ is a maximum such that $d_e\geq \textrm{Second}(S)$)}
  \STATE label Second($S$) with $\delta(d,S)$
  \STATE remove Second($S$) from stack $S$
 \ENDIF
 \STATE stack $e$ to $S$
 \ENDFOR
 \WHILE{length of $S$ $>2$}
 \STATE  label First($S$) with $\delta(d,S)$
 \STATE pop stack $S$
 \ENDWHILE
 \STATE label First($S$) and Second($S$) with $\delta(d,S)$
 \end{algorithmic}
\caption[Algorithm computing a scale labelling.]{\label{algo:computedelta}Algorithm to compute the scale labelling in $O(n)$ time. 
}
\end{algorithm}

\shortpaper{Once the labelling is complete, we find $K+2$ extrema having largest scale in time $O(n K)$ using $O(K)$ memory, then we remove all extrema having the same scale as the smallest scale in these $K+2$ extrema (removing at least one), we replace the first and the last extrema by $0$ and $n-1$ respectively. The result is an optimal segmentation having at most $K$ segments.}
\longpaper{
 Algorithm~\ref{algo:sortdelta} shows how once the labelling is complete, one can compute the segmentation in time $O(n K)$ which we consider to be linear time since $K$ is small compared to $n$. It also uses a fixed amount of memory ($O(K)$). If the segmentation points need to be sorted, then the complexity is $O(n K+ K \log K)$.
}
\longpaper{
\begin{algorithm}

 \begin{algorithmic}
 \STATE \textbf{INPUT:} an array $d$ containing the $y$ values indexed from $1$ to $n$
 \STATE \textbf{INPUT:} $K$ a bound on the number of segments desired
 \STATE \textbf{OUTPUT:} unsorted segmentation points (a $\delta$-segmentation)
 \STATE  $L \leftarrow $ empty array (capacity $K+3$)
 \FOR {$e$ is index of an extremum in $d$ having scale $\delta$, $e$ are visited in increasing order}
   \STATE insert $(e,\delta)$ in $L$ so that $L$ is sorted by scale in decreasing order (sort on $\delta$) using binary search
   \IF{length of $L$ is $K+3$}
     \STATE pop last(L)
   \ENDIF
 \ENDFOR
 \STATE remove all elements of $L$ having the scale of last(L)
 \STATE \textbf{RETURN:} the indexes in $L$ replacing first one by $1$ and last one by $n$
 \end{algorithmic}
\caption[Algorithm returning a segmentation using at most $K$ segments.]{\label{algo:sortdelta}Given the scale labelling, this algorithm will return a segmentation using at most $K$ segments. It is assumed that there are at least $K+1$ extrema to begin with.}
\end{algorithm}
}
\section{Experimental Results and Comparison to Top-Down Linear Spline}\label{experimental}

We compare our optimal $O(n K)$ algorithm with 
 the top-down linear spline algorithm~\cite{keogh04} which runs in $O(n K^2)$ time. It
successively segments the data starting with only one segment, each time picking the segment with the worse
linear regression error and finding the best segmentation point; the linear regression is not continuous from one segment to the other. The regression error can be computed in
constant time if one has precomputed the range moments~\cite{LemireCASCON2002}. We run through the
segments and aggregate consecutive segments having the same sign where the sign of a segment $[y_k,y_{k+1}]$ is defined by  $F(y_{k+1})-F(y_{k})$\longpaper{, setting 0 to be a positive sign (increasing monotonicity)}.

\subsection{Data Source}

 \longpaper{ECGs have a well known monotonicity structure with 5 commonly identifiable extrema per pulse (reference points P, Q, R, S, and T)  (see Fig.~\ref{pqrst}) though not all points can be easily identified on all pulses and the exact morphology can vary.} We used samples from the MIT-BIH Arrhythmia Database~\cite{PhysioNet}. \cutforlackofspace{We only present our results over one sample since we found that results did not vary much between data samples.} These ECG recordings used a sampling rate of 360 samples per second per channel with 11-bit resolution\longpaper{ (see Fig.~\ref{basicecg})}. We keep 4000 samples (11 seconds) and about 14 pulses, and we do no preprocessing such as baseline correction. We can estimate that a typical pulse has about 5 ``easily'' identifiable monotonic segments. Hence, out of 14 pulses, we can estimate that there are about 70 significant monotonic segments, some of which match the domain-specific markers (reference points P, Q, R, S, and T). A qualitative description of such data is useful for pattern matching applications.
 
\longpaper{
\begin{figure}
    \center{
    \includegraphics[width=0.8\columnwidth]{pqrst}
    }

\caption{\label{pqrst}Schema of an ECG pulse with commonly identified reference points (PQRST).}
\end{figure}
}
\longpaper{
\begin{figure}
    \center{
    \includegraphics[width=0.99\columnwidth]{100_original}
    }

\caption{\label{basicecg}ECG recordings from the MIT-BIH Arrhythmia Database.}
\end{figure}
}

\subsection{Results}

We implemented both the scale-based segmentation algorithm and the $L_2$ norm top-down linear spline approximation algorithm in
Python(version 2.3)
. 
Each run was repeated 3 times and \shortpaper{we observed that the scale-based segmentation implementation is faster than the top-down linear
spline approximation implementation by a factor of 10.}
\longpaper{the
average time is presented in Fig.~\ref{timevsk}. We chose a single representative sample: results do not vary significantly. The scale-based segmentation implementation is faster than the top-down linear
spline approximation implementation and the gains become considerable as $K$ increases.}

\longpaper{
\begin{figure}
\centering\includegraphics[width=0.8\columnwidth]{100_TimevsK}
\caption{\label{timevsk}Time vs. number of segments $K$.}
\end{figure}
}

\begin{figure}
\centering\includegraphics[width=1.00\columnwidth]{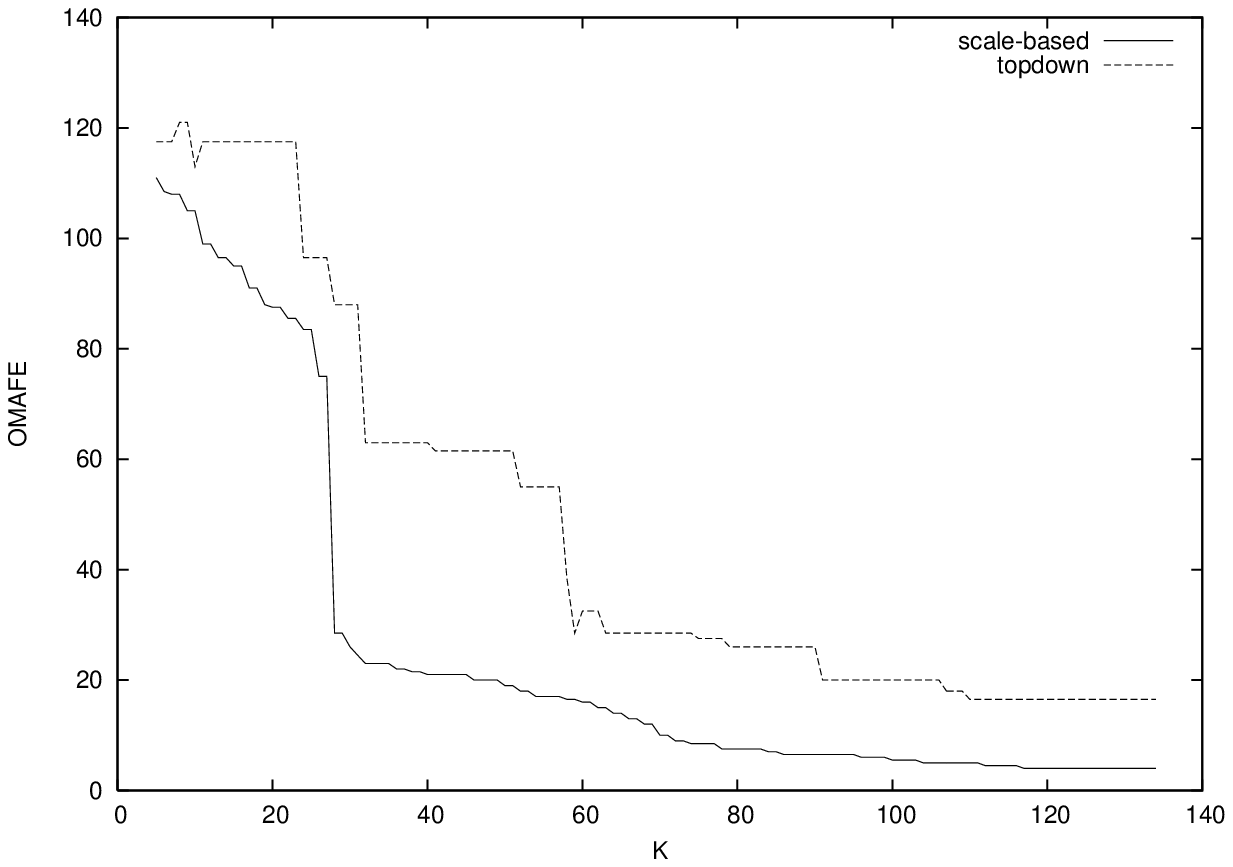}

\caption{\label{omafevsk}Results from experiments over ECG data: the optimal algorithm is considerably more accurate.}
\end{figure}


\longpaper{We want to
determine how well the top-down linear spline algorithm does comparatively. OMAFE is an absolute and not
relative error measure, but because the range
of the ECGs under consideration is roughly between 950 and 1150, we expect the OMAFE to never exceed 100 by much. }
The OMAFE with respect to the maximal number of segments ($K$) is given in Fig.~\ref{omafevsk}\longpaper{: it is a ``monotonicity spectrum.''}\shortpaper{.}
By counting on about 5 monotonic segments per pulse with a total of 14 pulses, there should about 70 monotonic segments in the 4000 samples under consideration.   We see that the decrease in OMAFE with the addition of new segments starts to level off between 50 and 70 segments as predicted. \longpaper{The addition of new segments past 70 ($K>70$) has little impact. The scale-based algorithm is optimal, but also at least 3 times more accurate than the top-down algorithm for larger $K$ and this is consistent over other data sets. In fact, the OMAFE becomes practically zero for $K>80$ whereas the OMAFE of the top-down linear regression algorithm remains at around 20, which is still significant. 
OMAFE of the scale-based algorithm is a non increasing function of $K$, a consequence of optimality.} 


\longpaper{\section{Future work}
Future work will focus on choosing the optimal number of segments for given applications.
We also plan to investigate the applications of the monotonicity spectrum as a robust
analysis.}

\bibliographystyle{latex8}
\bibliography{../MBR/modelabstraction,../MBR/yuhong,metric1d} 

\begin{thebibliography}{1}\setlength{\itemsep}{-1ex}\small

\bibitem{PhysioNet}
{A. L. Goldberger \textit{et al.}}
\newblock {PhysioBank, PhysioToolkit, and PhysioNet}.
\newblock {\em Circulation}, 101(23):215--220, 2000.

\bibitem{YLBIJCAI05}
M.~Brooks, Y.~Yan, and D.~Lemire.
\newblock Scale-based monotonicity analysis in qualitative modelling with flat
  segments.
\newblock In {\em IJCAI'05}, 2005.

\bibitem{LemireCASCON2002}
D.~Lemire.
\newblock Wavelet-based relative prefix sum methods for range sum queries in
  data cubes.
\newblock In {\em CASCON}. IBM, October 2002.

\bibitem{keogh04}
T.~Palpanas, M.~Vlachos, E.~Keogh, D.~Gunopulos, and W.~Truppel.
\newblock Online amnesic algorithm of streaming time series.
\newblock In {\em {ICDE}}, 2004.

\bibitem{Ubhaya1974}
V.~A. Ubhaya.
\newblock Isotone optimization {I}.
\newblock {\em Approx. Theory}, 12:146--159, 1974.

\end{thebibliography}

\end{document}